\documentclass{amsart}
\usepackage{amsmath, euscript, amssymb, amsfonts}

\theoremstyle{definition}

\theoremstyle{remark}

\newcommand{\oR}{{\mathbb R}}

\newcommand{\oC}{{\mathbb C}}

\newcommand{\oS}{{\mathbb S}}

\def\d{{\mathrm d}}
 \newcommand{\pr}{\mathop{\rm pr}\nolimits}

\begin{document}

\title{Carrier cones of analytic functionals}

\author{M.~A.~Soloviev}
\address{I.~E.~Tamm Theory Department, P.~N.~Lebedev
Physical Institute, Leninsky prospect 53, Moscow 119991, Russia}
\email{soloviev@lpi.ru}
\thanks{This work was supported in part by the grants
RFBR-05-01-01049 and LSS-1578.2003.2}

 \begin{abstract}
 We prove that every continuous linear functional on the space
  $S^0(\oR^d)$ consisting of the entire analytic functions whose
   Fourier transforms belong to the Schwartz space
   $\mathcal D$ has a unique minimal carrier cone in $\oR^d$, which
  substitutes for the support. The proof is based on a relevant
  decomposition theorem for  elements of the spaces $S^0(K)$
   associated naturally with closed cones $K\subset \oR^d$.
   These results, essential for
  applications to nonlocal quantum field theory, are similar to
  those obtained previously for functionals on the Gelfand-Shilov
  spaces $S^0_\alpha$, but their derivation is more sophisticated because
  $S^0(K)$ are not DFS spaces and have more complicated
  topological structure.
  \end{abstract}

\hfill Preprint FIAN/TD/13-05

\bigskip

\hfill math-ph/0507011

\vspace{2cm}

\maketitle

    \section{Introduction}
    The  Gelfand-Shilov test-function spaces $S^\beta$ and $S^\beta_\alpha$
    with the index  $\beta<1$, consisting of entire analytic functions~\cite{GS},
    find an   interesting application in quantum field theory (QFT), where
    they are used as a functional domain of definition of nonlocal
    fields.
     This permits the extension of
      the basic results of the Wightman axiomatic approach~\cite{SW}
       to nonlocal interactions, see~\cite{S4} and  references therein.
       The spaces $S^0$ and $S^0_\alpha$ play a dominant role in
      these applications  because the Fourier transforms of their elements
       have compact supports and these spaces are hence suitable
       for fields with      arbitrarily singular high-energy behavior.
       The central problem in constructing nonlocal QFT is an
      adequate generalization of the microcausality condition. The
      most natural formulation is attained through the use
      of the notion of a carrier cone (quasi-support) of
       non-localizable generalized functions. The
      existence of a unique minimal carrier cone for every
      generalized function defined on $S^0_\alpha$
      has been proved in~\cite{S0} with the aid of techniques of the theory
      of hyperfunctions.  The proof  relied essentially on the nice
      topological properties of the spaces $S^0_\alpha$ which belong to the
      well-studied class~\cite{K} of DFS spaces.

The space $S^0$ is of particular interest because it coincides
with the Fourier transform of Schwartz's space  $\mathcal D$ of
infinitely differentiable functions of compact support. Initially,
this space was introduced just in the theory of Fourier
transformation of distributions and, in this context, it was
denoted by $Z$ in~\cite{GS}. The notation $S^0$ used in physical
literature emphasizes that this space is smallest among the spaces
 $S^\beta$. Furthermore, the original definition of $S^\beta$ was
formulated in terms of real variables  and our concern here
        is precisely with carriers in the real space $\oR^d$, although
         the representation  in terms
        of complex variables will be used as an efficient method
        of the analysis. As noted in~\cite{GS}, the continuous linear
         functionals defined on $Z$
       and composing the dual space $Z'$ are analytic in the sense
       that their expansions in  Taylor series are convergent in the strong
       topology of $Z'$.
       The proof of the existence of a quasi-support for every functional
       of the class $Z'=S^{\prime\, 0}$ has some subtleties
        as compared to $S^{\prime\,0}_\alpha$, which  are elucidated in this
       paper. It should be noted that the arguments given below
        admit the extension of the results
       to multilinear forms, which is important for the applications
       to nonlocal QFT and to noncommutative field theory, where  such test
       functions also came recently into use~\cite{FP}.

Our construction is based on exploiting spaces related to $S^0$
and associated with open and closed cones in $\oR^d$. In Sec.~2,
we recall their definition introduced in~\cite{FS} and describe
the topological properties of these spaces. The existence of a
smallest carrier cone for every generalized function on $S^0$ is
deduced from a decomposition theorem which asserts that each
element of the space $S^0(K_1\cap K_2)$ associated with the
intersection of two closed cones can be decomposed into a sum of
functions belonging to $S^0(K_1)$ and $S^0(K_2)$. There are three
steps in proving this theorem. In Sec.~3, we first perform a
 decomposition into smooth functions satisfying the required
 restrictions on the behavior at infinity. The analyticity is
 restored in Sec.~5 with the use of
  H\"ormander's classical results~\cite{H2}  concerning  solutions
  of the nonhomogeneous Cauchy-Riemann equations. However, H\"ormander's
   $L^2$-estimates involve plurisubharmonic functions, whereas
   the indicator functions defining  $S^0(K)$
    are not of such a form. Because  of this, we run into the problem of
approximating the indicator functions by  plurisubharmonic ones,
which is solved  in Sec.~4. The main theorem
   on carrier cones is proved in Sec.~6. In concluding Sec.~7, we
   point some possible applications of the obtained results.

\section{Spaces over cones and their topology}

   Let $U$ be an open cone in $\oR^d$.
The space  $S^{0,b}(U)$ is defined to be the intersection
(projective limit) of the family of Hilbert spaces $H^{0,B}_N(U)$,
$B>b$, $N=0,1,2,\dots$,  consisting of entire analytic functions
on $\oC^d$ and endowed with the scalar products
  \begin{equation} {\langle f,\,g\rangle}_{U,B,N} = \int
\overline{f(z)} g(z)\,(1+|x|)^{2N}\,e^{-2Bd(x,U) - 2B|y|}\, {\rm
d}\lambda, \qquad z=x+iy,
   \label{*}
   \end{equation} where
$d(x,U)=\inf_{\xi\in U}|x-\xi|$ is the distance of $x$ from $U$
and ${\rm d}\lambda={\rm d}x{\rm d}y$ is the Lebesgue measure
  on $\oC^d$. Hereafter we assume the norm $|\cdot|$ on  $\oR^d$
  to be Euclidean.  It is easily seen from
  Cauchy's integral formula that the system of norms
 $\|f\|_{U, B,N}$ determined by scalar products~\eqref{*}
is equivalent to the system $\|f\|'_{U, B, N}=
\sup_z|f(z)|e^{-\rho_{U,B,N}(z)}$, where
   \begin{equation}
\rho_{U,B,N}(z)=-N\ln(1+|x|)+Bd(x,U) +B|y|.
    \label{rho}
   \end{equation}
In particular, if $U=\oR^d$, then $d(x,\oR^d)=0$ and
$S^{0,b}(\oR^d)$
   coincides with the space
   $Z(b)$ which is the Fourier transform of the space of
   all   smooth functions with support in the closed ball of radius
   $b$, see~\cite{GS}.
   However, the employment of Hilbert norms is  preferable
    in some cases, including the  proofs given below.
 The space $S^0(U)$ is defined to be the union (inductive limit)
  of the family $S^{0,b}(U)$, $b>0$. Clearly, there are natural
  inclusion maps  $S^0(U_1)\to S^0(U_2)$ for $U_1\supset U_2$.
  As stated above, the Fourier operator transforms $S^0(\oR^d)$
into the Schwartz space $\mathcal D(\oR^d)$.

 Let $v$ be a   continuous linear functional on  $S^0(\oR^d)$.
  We say that $v$ is carried by a closed cone $K\subset \oR^d$ if
  this functional has a continuous extension to every space $S^0(U)$, where
   $U\Supset K$.\footnote{For arbitrary cones $V_1$, $V_2$, the
notation $V_1\Subset V_2$ means that $\Bar V_1\setminus
\{0\}\subset V_2$, where $\Bar V_1$ is the closure of $V_1$.} This
extension, if it exists, is unique because $S^0(\oR^d)$ is dense
in each $S^0(U)$, see~\cite{S2,S1}. Therefore  the stated property
implies that $v$ extends to the space
 \begin{equation}
 S^0(K)=\varinjlim_{U\Supset K}S^0(U),
 \label{**}
   \end{equation}
   and the set of  functionals  carried by
  $K$ is algebraically identified with the dual space of $S^0(K)$.
 It is worth noting that in the case of  degenerate closed cone
 consisting of the origin,
the set of open cones $U$ such that $U\Supset \{0\}$ is not
directed. For the sake of unification, the cone $\{0\}$ can be
added to the totality of open cones with   its
   associated space $S^0(\{0\})$ defined as the space
   of all entire analytic   functions of exponential type.
   This stipulation is quite natural, because really we are
    dealing here with
   a sheaf of spaces over  $(d-1)$-dimensional sphere $\oS^{d-1}$
    compactifying
   $\oR^d$ and the cone $\{0\}$ corresponds to the empty open set
   in $\oS^{d-1}$. For clearness, we prefer to say about cones
   instead of subsets  of  $\oS^{d-1}$, bearing physical applications
   in mind. All the spaces $S^0(U)$ are embedded in $S^0(\{0\})$
and the decomposition theorem derived below shows that the
inductive topology on $S^0(\{0\})$ determined by the injections
  $S^0(U)\to S^0(\{0\})$, where $U\ne \{0\}$,  coincides with the
  topology defined on the basis of~\eqref{*} with $d(x,0)=|x|$. Thus
     definition~\eqref{**} is applicable to $\{0\}$ even without
     this stipulation but we find it to be convenient.

The spaces $S^{0,b}(U)$ belong to the class FN of nuclear
Fr\'echet spaces. This fact, which is  essential for applications
to QFT, can be established in a manner analogous to that used
in~\cite{S1} for $S^{0,b}_{\alpha,a}(U)$. As a consequence,
$S^{0,b}(U)$ also belong  to the Fr\'echet-Schwartz class FS and
are  Montel spaces and hence reflexive. The spaces $S^0(U)$ and
$S^0(K)$, being
 inductive limits of  sequences of such spaces, inherit
nuclearity (see~\cite{Sch}) and are obviously Hausdorff spaces.
The spaces $S^0(U)$ are complete, as is proved in~\cite{S3} with
the use of acyclicity of the injective sequence $S^{0,\nu}(U)$,
$\nu=1,2,\dots$.  Together with nuclearity and barrelledness, this
implies that they are also Montel spaces, see Exer.~19 in Chap.~IV
in~\cite{Sch}. Whether $S^0(K)$ have such properties is still an
open question, but their completions have them because nuclearity
and barrelledness are preserved after completion. However the
spaces $S^0(U)$ and $S^0(K)$ are nonmetrizable. This fact is well
known for $S^0(\oR^d)$ and follows, e.g., from Proposition 5 in
Sec. VII in~\cite{RR}, because  $\mathcal D(\oR^d)$ is a strict
inductive limit. In the case $U\ne\oR^d$, one can use Corollary
7.3 in~\cite{P}, where injective limits of acyclic sequences are
considered.

\medskip
\noindent
 {\bf Proposition 1.} {\it The spaces $S^0(U)$ and
$S^0(K)$ are not DFS spaces, with one exception $S^0(\{0\})$.

\medskip
\noindent
 Proof.} It suffices to show that the dual spaces
$S^{\prime\, 0}(U)$, $S^{\prime\, 0}(K)$ endowed with the strong
topology are nonmetrizable because the class DFS consists of the
strong dual spaces of FS spaces and each DFS space is reflexive.
For this purpose we use Theorem~A
 in~\cite{G}. In the terminology of~\cite{G}, a locally convex $E$
 space is called an $\mathcal{LF}$ space if it is a Hausdorff space
 representable as the inductive limit of a sequence of Fr\'echet
 spaces $E_j$ (without the  more usual assumption~\cite{Sch} that
 this limit is strict).  Grothendieck's theorem~\cite{G} asserts that
 if $u$ is a continuous linear mapping of a Fr\'echet space $F$ to such an
$\mathcal{LF}$ space $E$, then $u(F)$ is contained in one of the
spaces $E_j$. In the case under consideration, $S^{\prime\,
0,b}(U)$ is the inductive limit of the sequence of strong dual
spaces of the Hilbert spaces $H^{0, b+1/N}_N(U)$ and so has the
structure of $\mathcal{LF}$ space. If $S^{\prime\, 0}(U)$ (or
$S^{\prime\, 0}(K)$) were a Fr\'echet space, then its image under
the natural injection to $S^{\prime\, 0,b}(U)$ would be contained
in the dual space of an $H^{0, b+1/N}_N(U)$ but such is not the
case except for $S^{\prime\,0}(\{0\})$, when the factor
$(1+|x|)^{2N}$ in ~\eqref{*} can be dropped out. This completes
the proof.

\medskip
\noindent
 {\it Remark} 1. By the same argument, none of the spaces
$S^\beta(\oR^d)$, $\beta\ge 0$, is a DFS space.

 \section{A smooth decomposition}

Our objective is to prove the following theorem.

\medskip
\noindent
 {\bf Theorem 1}. {\it  If $U_1$, $U_2$, and $U$ are open cones
 in $\oR^d$ such that $\bar U_1\cap \bar U_2\Subset U$ and
 $f\in S^0(U)$, then $f=f_1+ f_2$ with
 $f_i\in S^0(U\cup U_i)$, $i=1,2$.}

\medskip
  Using the dilation invariance of the spaces involved, we assume that
   $\|f\|_{U, 1, N}<\infty$ without loss of generality. As noted
   above, we first decompose $f$ into smooth functions which
   behave at infinity as elements of   $S^0(U\cup U_i)$ and
   subsequently restore  the analyticity. The smooth decomposition
    can be  performed with the help of the simple geometrical lemma.

\medskip
\noindent
 {\bf Lemma 1}. {\it For every triple of open cones
 $U_1$, $U_2$, $U$ satisfying $\Bar U_1 \cap\Bar U_2\Subset
U$, there is a smooth function $\chi(x)\ge 0$ with the properties:
\begin{gather}
|\chi(x)|e^{d(x,U)}\le Ce^{bd(x,U\cup U_1)}, \label{1}\\
|1-\chi(x)|e^{d(x,U)}\le Ce^{bd(x,U\cup U_2)}, \label{2}\\
\left|\frac{\partial\chi}{\partial x_j}\right|e^{d(x,U)}\le
Ce^{bd(x,U\cup U_1\cup U_2)},\quad j=1,\dots, d,
 \label{3}
\end{gather}
where $C$ and $b$ are positive constants.}

\medskip
Using such a function, we set
 $$
 f=f_1+f_2,\quad f_1(z)=f(z)\chi(x),
 \quad f_2(z)=f(z)(1-\chi(x))
   $$
  and then~\eqref{1} and~\eqref{2} (with $b\ge 1$) result in the
  iequalities
 \begin{equation}
 \|f_1\|_{U\cup U_1,b,N}\le C\|f\|_{U, 1, N},\quad
 \|f_2\|_{U\cup U_2,b,N}\le C\|f\|_{U, 1, N},\qquad N=0,1,2,\dots
  \label{4}
  \end{equation}

\medskip
\noindent
   {\it   Proof of Lemma 1.} We recall that the intersection of a cone $V$
   with the unit sphere is named its projection and denoted by
     $\pr V$. The condition $\Bar
U_1 \cap\Bar U_2\Subset U$ implies that the closed cones $V_1=\Bar
U_1\setminus U$ and $V_2=\Bar U_2\setminus U$ have disjoint
projections. Therefore the distance of $\pr V_1$ from $V_2$, as
well as that of $\pr V_2$ from $V_1$, is positive. In the case of
Euclidean metric, these two distances are equal. In fact, if the
first one is attained at the points  $x_1\in \pr V_1$, $x_2\in
V_2$, then the relation $|x_1-x_2|^2=|x_1|x_2|-x_2/|x_2||^2$ shows
that $d(\pr V_1, V_2)\ge d(\pr V_2, V_1)$, and the reverse is also
true by symmetry. This quantity will be referred to as the angular
separation between   $V_1$ è $V_2$ and denoted by $\theta$.

 Let us introduce the auxiliary open cone
$$
 W=\left\{\xi\in \oR^d\colon d(\xi, V_2)< \frac{\theta}{2}|\xi|\right\}
 $$
 and let $\chi_0$ be a nonnegative function of the class
$C^\infty$ with support in the unit ball and whose integral is
unity. We define $\chi(x)$ by
\begin{equation}
 \chi(x)=\int\limits_W\!\chi_0(x-\xi)\,{\rm d}\xi .
 \notag
  \end{equation}
  and claim that this function satisfies
  inequalities~\eqref{1}--\eqref{3} with
   $b$  arbitrarily close to $2/\theta$.
 To demonstrate this, we use further two cones $W_1$, $W_2$
specified in a similar way to  $W$ but with parameters $\theta_1$,
$\theta_2$ subject to the conditions $\theta<\theta_1<2\theta$ and
$0<\theta_2<\theta$. It is easy to see that
\begin{equation}
 d(x, V_1)\ge \left(\theta-\frac{\theta_1}{2}\right )|x|\quad
 \mbox{for all}\quad x\in W_1 .
\label{6}
  \end{equation}
Indeed, if this is not the case, then $d(\pr W_1, V_1)<
\theta-\theta_1/2$ and there are points $x_1\in \pr V_1$, $x\in
W_1$ such that $d(x_1,x)< \theta-\theta_1/2$. Besides, $|x|<1$
because the distance of the unit vector $x_1$ to $W_1$ is attained
inside the unit ball. There is also a point $x_2\in V_2$ such that
$d(x, x_2)<\theta_1|x|/2<\theta_1/2$ and  the triangle inequality
gives $d(x_1,x_2)<\theta$, which contradicts the definition of
$\theta$. The support of $\chi$ is contained in the 1-neighborhood
of $W$ and hence in the union of $W_1$ with a ball of sufficiently
large radius $R$.  The condition~\eqref{1} is trivially satisfied
inside the ball with $C=e^R$ and any $b$. The inequality~\eqref{6}
ensures the fulfilment of~\eqref{1} in $W_1$ with $C=1$ and
$b=(\theta-\theta_1/2)^{-1}$ because $d(x, U)\le |x|$ and $d(x,
U\cup U_1)=\min\{d(x,U), d(x, U_1\setminus U)\}$, where
$U_1\setminus U\subset V_1$. Further, $1-\chi(x)=0$ for  all the
points $x\in W$ whose distance from the boundary of $W$ is greater
than 1 and, in particular, for the points of $W_2$ outside a
sufficiently large ball. Considering that
\begin{equation}
 d(x,V_2)\ge\frac{\theta_2}{2}|x|\quad
 \mbox{for all}\quad x\notin W_2,
 \label{7}
  \end{equation}
we infer that~\eqref{2} is satisfied with $b=2/\theta_2$. The
derivatives of $\chi$  are uniformly bounded and their supports
are contained in the 1-neighborhood of the boundary of $W$.
Therefore both of inequalities~\eqref{6}, \eqref{7} are satisfied
at the points of these supports beyond a sufficiently large ball.
Consequently,~\eqref{3} also holds with any $b>2/\theta$ as
$\theta_1$ and $\theta_2$ can be taken arbitrarily close to
$\theta$. Lemma~1 is thus proved.

\section{Going to plurisubharmonic functions}

We go on to prove Theorem 1. To obtain an analytic decomposition,
we write
 \begin{equation}
 f=f'_1+f'_2,\quad  f'_1=f_1-\psi,\quad f'_2=f_2+\psi
 \notag
\end{equation}
and subject $\psi$  to the equations
\begin{equation}
\frac{\partial\psi}{\partial\bar z_j}=\eta_j ,
 \label{8}
\end{equation}
where
\begin{equation}
\eta_j\stackrel{{\rm def}}{=}f\frac{\partial\chi}{\partial\bar
z_j} =\frac{1}{2}f\frac{\partial\chi}{\partial x_j},\qquad
j=1,\dots,d.
 \label{9}
\end{equation}
    The functions  $\eta_j(z)$ satisfy the estimate
\begin{equation}
 |\eta_j(z)|\le C_{j,N} \|f\|_{U, 1, N}\,e^{\rho_{{}_{U\cup U_1\cup U_2,
b,N}}(z)},
 \label{10}
 \end{equation}
 where notation~\eqref{rho} is used.
In fact, from Cauchy's integral formula we have
\begin{equation}
|f(z)|\le C\|f\|_{L^2(\mathcal B)},
 \notag
  \end{equation}
  where $\mathcal B$ is any bounded neighborhood of the point
   $z$ in $\oC^d$.  Taking the unit ball for $\mathcal B$
    and applying the triangle inequality to every item
   on the right-hand side of~\eqref{rho}, we infer that
      $-\rho_{U,B,N}(z)\le  -\rho_{U,B,N}(\zeta)+N\ln 2+2B$
      for all $\zeta\in \mathcal B$ .
   Therefore
\begin{equation}
|f(z)|^2 e^{-2\rho_{{}_{U, 1,N}}(z)}\le C_N\int_{\mathcal
B}|f(\zeta)|^2 e^{-2\rho_{{}_{U, 1,N}}(\zeta)}\d \lambda \le C_N
\|f\|^2_{U, 1, N}
 \notag
  \end{equation}
and, coupled with~\eqref{3}, this gives~\eqref{10}. It remains to
show that there exist a solution of Eqs.~\eqref{8} with the
required behavior at infinity.  For this purpose we will use
H\"ormander's $L^2$-estimates but first prove another lemma.

\medskip
\noindent
 {\bf Lemma 2.} {\it Let $V$ be an open cone in $\oR^d$
and $B>2edb$. For each function $\eta(z)$, $z\in \oC^n$,
satisfying the estimate
\begin{equation}
 |\eta(z)|\le C_N\,e^{\rho_{{}_{V, b,N}}(z)}
 \label{11}
 \end{equation}
 there is a plurisubharmonic function $\varrho(z)$ with values in
  $(-\infty,+\infty)$ such that}
\begin{equation}
 |\eta(z)|\leq e^{\varrho(z)}\leq
 C'_N\,e^{\rho_{{}_{V, B,N}}(z)}.
 \label{12}
 \end{equation}

\medskip
\noindent
   {\it   Proof.} For brevity, we set $b=1$.  As shown in~\cite{S1},
for each $\alpha>1$ and  $\sigma>2$, there is a sequence of
functions $\varphi_n\in S^0_\alpha(\oR)$, $n=0,1,2,\dots$, with
the properties:
 \begin{align}
&|\varphi_n(z)| \le C\exp\{\sigma|y|-|x|^{1/\alpha}\},
\label{c1}\\
&\ln|\varphi_n(iy)|\ge |y|, \label{c2}\\
&\ln|\varphi_n(z)| \le \sigma|y|-n\ln^+(|x|/n)+A, \label{c3}
 \end{align}
where $\ln^+r=\max (0,\ln r)$ and the constants $C$,  $A$ are
independent of $n$.~\footnote{Here we mean that $n\ln^+(|x|/n)=0$
for $n=0$.}

This sequence is the main tool of our proof and we also use the
auxiliary function
\begin{equation}
 H(\xi)=\sup_y\{\ln|\eta(\xi+iy)|-|y|\}.
\label{c4}
 \end{equation}
By condition~\eqref{11} it satisfies the inequality
\begin{equation}
 H(\xi)\le \ln C_N-N\ln(1+|\xi|)+d(\xi,V).
 \label{c5}
 \end{equation}
First we consider the simplest one-dimensional case $V=\oR_-$,
$d(\xi,V)=\theta(\xi)\,|\xi|$, where $\theta(x)$ is Heaviside's
step-function. Let $\Phi_n(z)=\ln|\varphi_n(ez)|$. The function
$\Phi_n$ is subharmonic according to~\cite{V}, \S\,II.9.12. As a
candidate for the desired function $\varrho$, we take the upper
envelope of the family  $\Phi_n(z-\xi)+H(\xi)$, allowing the index
$n$ to depend on the point $\xi$ running $\oR$. The functions of
this family are locally uniformly bounded from above and hence
their upper envelope is also subharmonic (ibid, \S\,II.9.6).
Furthermore, it obviously dominates $\ln|\eta(z)|$
because~\eqref{c2} and \eqref{c4} imply
\begin{equation}
\sup_\xi\{\Phi_n(z-\xi)+H(\xi)\}\ge \Phi_n(iy)+H(x)\ge
\ln|\eta(z)|.
 \label{c6}
 \end{equation}
Let us show that  $n(\xi)$ can be chosen so that to ensure the
second of inequalities~\eqref{12}. For $\xi<0$, when
$d(\xi,\oR_-)=0$, there is no problem and we may set $n(\xi)=0$
because~\eqref{c1}, together with the elementary inequalities
\begin{equation}
-|x-\xi|^{1/\alpha}\le \tilde C_N-N\ln(1+|x-\xi|),\quad
\ln(1+|x-\xi|)+ \ln(1+|\xi|)\ge \ln(1+|x|),
 \label{c7}
 \end{equation}
 implies that
$$
\sup_{\xi<0}\{\Phi_0(z-\xi)+H(\xi)\}\le\Tilde C'_N+e\sigma|y|
-N\ln(1+|x|).
 $$
By using~\eqref{c7}, we also obtain the estimate
\begin{equation}
\beta\,\Phi_0(z-\xi)-N\ln(1+|\xi|)\le\Tilde C''_N+\beta e\sigma|y|
-N\ln(1+|x|),
 \label{c8}
 \end{equation}
 which holds for any $\beta>0$ and for all $\xi$ and shows
  that  difficulties emerge only from the linear growth
 of the term $d(\xi, V)$ in~\eqref{c5}.

 Let
$\xi\ge 0$ and so $d(\xi,\oR_-)=\xi$. Let, in addition, $e|x-\xi|>
n$. Then
\begin{equation}
n\ln(e|x-\xi|/n) +e d(x,\oR_-)\ge n\ln(e\xi/n).
\label{c9}
 \end{equation}
This is obvious for  $|x-\xi|>\xi$. For the other points, it
suffices to use the inequality $\theta(x)\,x\ge\xi-|x-\xi|$ and
take into account that the function $n\ln(\lambda/n)-\lambda$ is
monotone decreasing in the interval $[n,e\xi]$.
Applying~\eqref{c3} and \eqref{c9}, we get
$$
\Phi_n(z-\xi)+\xi \le A+\sigma e |y|+ed(x,\oR_-)-n\ln(e\xi/n)+\xi.
 $$
 We take $n(\xi)$ to be the integral part of the number $\xi$.
 Then
$n\ln(e\xi/n)\ge n>\xi-1$ and
\begin{equation}
\Phi_{n(\xi)}(z-\xi)+\xi \le \tilde A+\sigma e |y|+ed(x,\oR_-).
 \label{c10}
 \end{equation}
 An analogous inequality holds for  $e|x-\xi|\le n$,
when $\ln^+(e|x-\xi|/n)=0$. Indeed, then $\xi\le
\theta(x)\,|x|+|x-\xi|\le \theta(x)\,|x|+\xi/e$ and hence $\xi\le
ed(x,\oR_-)$. Thus, inequality~\eqref{c10} (with a proper
constant) is valid for all $\xi\ge 0$. In combination with
estimate~\eqref{c8}, where we set $\beta=B/(e\sigma)-1$, it shows
that the upper envelope
\begin{equation}
\varrho(z)=\varlimsup_{z'\to
z}\sup_\xi\{\beta\Phi_0(z'-\xi)+\Phi_{n(\xi)}(z'-\xi)+H(\xi)\}
\label{c11}
 \end{equation}
 satisfies all the requirements.~\footnote{Taking the upper limit ensures
the upper semicontinuity of the resulting function and enters into
the definition~\cite{V} of upper envelope.}

In the general case of several variables and arbitrary open cone
$V\subset\oR^d$, we set
$\Phi_n(z)=\sum_{j=1}^d\ln|\varphi_n(e\sqrt{d}\,z_j)|$. Then
inequality \eqref{c6} holds true and~\eqref{c8} is replaced by
\begin{equation}
 \beta\,\Phi_0(z-\xi)-N\ln(1+|\xi|)\le\Tilde
C'''_N+\beta e d\sigma|y| -N\ln(1+|x|),
 \label{c12}
 \end{equation}
because $\sum_{j=1}^d|y_j|\le\sqrt{d}\,|y|$. For $x\notin V$ and
$e|x-\xi|> n$, we have
\begin{equation}
\sum_{j=1}^d n\ln\left(\frac{e\sqrt{d}}{n}|x_j-\xi_j|\right) +e
d(x,V)\ge n\ln\left(\frac{e}{n}d(\xi, V)\right).
 \notag
 \end{equation}
 To make sure of this, it suffices to  consider that
$$
\sum_{j=1}^d \ln^+|x_j|\ge\ln^+\frac{|x|}{\sqrt{d}}\qquad
\mbox{and} \qquad d(\xi, V)=\inf_{\xi'\in V}|\xi'-\xi|\le
d(x,V)+|x-\xi|.
$$
This time we take $n(\xi)$ to be the integral part of $d(\xi,V)$
and then  \eqref{c10} is changed for the inequality
\begin{equation}
\Phi_{n(\xi)}(z-\xi)+d(\xi, V) \le \tilde A'+\sigma e d
|y|+ed(x,V),
 \label{c13}
 \end{equation}
which holds for all $x$. Combining~\eqref{c13} with~\eqref{c12},
we conclude that the plurisubharmonic function defined
by~\eqref{c11} with $\xi\in \oR^d$ and $\beta=B/(e\sigma d)-1$
satisfies the conditions~\eqref{12}. So we have proved Lemma 2.

\section{The use of $L^2$-estimates}

\noindent
 {\it End of the proof of Theorem 1.} We set
 $$
 \tilde \varrho(z)=2\varrho(z)+(d+1)\ln(1+|z|^2),
 $$
  where $\varrho$
 is the upper envelope of the plurisubharmonic functions assigned
 to  $\eta_j(z)$, $j=1,\dots,d$ by Lemma~2 with $V=U\cup U_1\cup U_2$.
   Inequalities  $|\eta_j(z)|\leq  e^{\varrho(z)}$ implies that
   the functions $\eta_j$  belong to $L^2(\oC^d, e^{-\tilde \varrho}{\rm
  d}\lambda)$. By definition~\eqref{9}, the compatibility conditions
  $\partial \eta_j/\partial \bar{z}_k =\partial\eta_k/\partial
  \bar{z}_j$ are fulfilled.
  Therefore we can apply Theorem 15.1.2 in~\cite{H2}, which shows that
  the system of equations~\eqref{8} has a solution $\psi$ such that
   \begin{equation}
   2\int|\psi|^2 e^{-\tilde \varrho}(1+|z|^2)^{-2}{\rm d}\lambda
  \le \int|\eta|^2 e^{-\tilde \varrho}{\rm d}\lambda.
 \label{13}
   \end{equation}
  From~\eqref{12} and \eqref{13}, it follows that
 \begin{equation}
  \psi \in L^2\left(\oC^d,  e^{-2\rho_{V,B,N-d-3}}{\rm d}\lambda\right)
    \notag
   \end{equation}
for each $N$. When coupled with~\eqref{4}, this gives
  $f'_1\in S^{0}(U\cup U_1)$ and $f'_2\in S^{0}(U\cup U_2)$,
   which completes the proof.

\medskip
\noindent
 {\it Remark} 2. If $\bar U_1\cap \bar U_2=\{0\}$, then
every element $f\in S^0(\{0\})$ allows the decomposition $f=f_1+
f_2$, where
 $f_i\in S^0(U_i)$, $i=1,2$. This follows from the above arguments with
$d(x,0)=|x|$  and is included in the formulation of Theorem 1
under the stipulation of Sec.~2 that the cone $\{0\}$  is  open in
the sense that its projection is open.

\section{The existence of smallest carrier cones}

\noindent
 {\bf Theorem 2}. {\it For every  continuous linear functional on
 $S^0(\oR^d)$, there exists a unique minimal  closed carrier cone.

\medskip
\noindent
 Proof.} By Theorem 1,
\begin{equation}
  S^0(K_1\cap K_2)= S^0(K_1)+ S^0(K_2),
 \label{14}
   \end{equation}
 for each pair of closed cones in $\oR^d$. Indeed, if
  $f\in S^0(U)$, where
$U\Supset K_1\cap K_2\ne\{0\}$, then there are $U_i$ satisfying
the condition $\Bar U_1 \cap\Bar U_2\Subset U$. In the trivial
 case
$K_1\cap K_2=\{0\}$, above Remark is applicable because then there
are open cones  $U_i$ such that $K_i\Subset U_i$ è $\bar U_1\cap
\bar U_2=\{0\}$.

We now show that~\eqref{14} leads to the following dual relation
for functionals
\begin{equation}
  S^{\prime\, 0}(K_1\cap K_2)= S^{\prime\, 0}(K_1)\cap S^{\prime\,
  0}(K_2),
 \label{15}
   \end{equation}
where all the spaces are considered as subspaces of $S^{\prime\,
0}(\oR^d)$. The nontrivial part of~\eqref{15} is the assertion
that if a functional $v\in S^{\prime\, 0}(\oR^d)$ is carried both
by  $K_1$ and by $K_2$, then $K_1\cap K_2$ is also its carrier
cone. Let $v_i$ be continuous extensions of $v$ to $S^0(K_i)$ and
let $f\in S^0(K_1\cap K_2)$. Using the decomposition $f=f_1+f_2$,
where $f_i\in S^0(K_i)$, we put  $\hat v(f)=v_1(f_1)+v_2(f_2)$ and
claim that this extension of $v$ to $S^0(K_1\cap K_2)$ is well
defined. Indeed, if $f=f'_1+f'_2$ is another decomposition, then
$$
f_1-f'_1=f'_2-f_2\in S^0(K_1)\cap S^0(K_2)= S^0(K_1\cup K_2)
$$
and $v_1(f_1-f'_1)=v_2(f'_2-f_2)$ because $S^0(\oR^d)$ is dense in
$S^0(K_1\cup K_2)$ by Theorem~2 in~\cite{S2}. The functional $\hat
v$ is obviously continuous under the inductive topology  $\mathcal
T$ determined by the injections $S^0(K_i)\to S^0(K_1\cap K_2)$,
$i=1,2$, and this topology coincides with the original topology
$\tau$ on $S^0(K_1\cap K_2)$ by the open mapping theorem. Indeed,
$\tau$ is not stronger than
  $\mathcal T$ merely by the definition of $\mathcal T$  and we
  can apply   Grothendieck's  version~\cite{G} of the open
  mapping theorem,  because  $(S^0(K_1\cap K_2), \tau)$,
   being the inductive limit of Fr\'echet spaces,
  belongs to the class  $(\beta)$ in the terminology
  of~\cite{G}~\footnote{The spaces of this class
  are also termed ultrabornological} and $(S^0(K), \mathcal T)$
  is an $\mathcal{LF}$ space. This is the case because either of
     $S^0(K_i)$  is an $\mathcal{LF}$ space and $\mathcal T$
     coincides with the quotient topology of the sum
    $S^0(K_1)\oplus S^0(K_2)$ modulo a closed subspace, see~\cite{RR}.

A formula analogous to~\eqref{15} holds for the intersection of
each finite family of closed cones and now the existence of the
smallest carrier cone for $v$ can be established by standard
compactness arguments. Indeed, let $K$ be the intersection  of all
its carrier cones  and let  $U$ be an open cone such that
$K\Subset U$. The projections of the cones complementary to these
carriers cover the compact set $\pr\complement U$ and we can
choose a finite subcovering $\pr\complement K_j$ of these open (in
the topology of the unit sphere) covering. Consequently, the
functional $v$ is continuous in the topology of $S^0(U)$ and $K$
is its carrier cone, which completes the proof.

\section{Conclusion}
 Theorem  2 provides a basis for the development of the theory
 of Fourier-Laplace transformation for the functional class
$S^{\prime\, 0}$ and, in particular, for the derivation of a
Paley-Wiener-Schwartz-type  theorem analogous to  Theorem 4
in~\cite{S1} established for $S^{\prime\,0}_\alpha$. It is worth
noting that
 the  proof in Sec.~6 uses only the decomposability of  $f\in S^0(U)$
  into a sum of elements of $ S^0(U_i)$ under the condition $\Bar U_1
\cap\Bar U_2\Subset U$. The more precise formulation given in
Theorem 1 becomes essential in studying the carrier  properties of
 multilinear forms on $S^0\times\dots\times S^0$. A
similar formulation  was introduced in~\cite{Sm} in the framework
of DFS spaces $S^0_\alpha$. The generalization to multilinear
forms is important for applications to QFT because the vacuum
expectation values of quantum fields are  such forms.
Specifically, this development is necessary to an understanding of
the interrelation between the asymptotic commutativity
condition~\cite{S4,S2}, which ensures the normal spin-statistics
connection and CPT invariance of nonlocal QFT, and the regularity
properties~\cite{St} of retarded Green's functions in momentum
space that are required for the construction of  scattering
theory. The relation between the carriers of multilinear forms
with respect to their arguments and the carrier cones  of those
functionals that are generated by these forms by the kernel
theorem is also essential in using the analytic test functions in
noncommutative field theory, where a causality
condition~\cite{Alv} is formulated with a light wedge instead of
the light cone.  These questions will be discussed at length  in a
subsequent paper.

\end{document}